\documentclass[12pt,epsf]{article}
\usepackage{graphicx}

\begin{document}

\sloppy
\begin{flushright}{SIT-HEP/TM-23}
\end{flushright}
\vskip 1.5 truecm
\centerline{\large{\bf Formation of monopoles and domain walls after
brane inflation}} 
\vskip .75 truecm
\centerline{\bf Tomohiro Matsuda
\footnote{matsuda@sit.ac.jp}}
\vskip .4 truecm
\centerline {\it Laboratory of Physics, Saitama Institute of
 Technology,}
\centerline {\it Fusaiji, Okabe-machi, Saitama 369-0293, 
Japan}
\vskip 1. truecm
\makeatletter
\@addtoreset{equation}{section}
\def\theequation{\thesection.\arabic{equation}}
\makeatother
\vskip 1. truecm

\begin{abstract}
\hspace*{\parindent}
We study cosmological defect formation that is induced by brane dynamics
 after brane inflation. 
The cosmological defects are corresponding to the branes that have less than
 three spacial dimensions in the uncompactified spacetime.
Contrary to the previous arguments, production of monopoles and
domain walls are not always negligible.
Monopoles and domain walls are formed by the branes extended between
 mother branes. 
\end{abstract}

\newpage
\section{Introduction}
Models with more than four dimensions are interesting, because all the
physical ingredients of the Universe might be unified in a higher
dimensional theory.
String theory is the most promising scenario where quantum
gravity is included by the requirement of additional dimensions and
supersymmetry.
The idea of large extra dimension\cite{Extra_1} is important,
because it might solve the hierarchy problem.
In this case, the observed Planck mass is obtained by the relation
$M_p^2=M^{n+2}_{*}V_n$, where $M_{*}$ and $V_n$ denote the fundamental
scale of gravity and the volume of the $n$-dimensional compact space,
respectively. 
In the scenarios of large extra dimension, fields in the standard model
are expected to be localized on a wall-like structure, while the graviton
propagates in the bulk. 
In the context of string theory, a natural embedding of
this picture is realized by brane construction.
The brane models are interesting from both phenomenogical and 
cosmological viewpoints.

Analyses of cosmological defect formation are important in brane models
as well as in usual cosmological
models.\footnote{Inflation in models of low fundamental scale 
is interesting\cite{low_inflation, matsuda_nontach,matsuda_defectinfla}.
Baryogenesis is discussed in ref.
\cite{low_baryo, Defect-baryo-largeextra, 
Defect-baryo-4D}, where defects play important roles.
Non-static brane configurations such as brane defects or brane
Q-balls\cite{BraneQball, matsuda_MQCD_defect} are important, because
we are expecting that future cosmological observation will reveal
the cosmological evolution of the Universe, which will also reveal
the physics beyond the standard model.
If one wants to know what kinds of brane defect are produced
in the early Universe, one needs to understand how they are formed.}
One may consider three types of brane defects.
\begin{itemize}
\item Defects are branes.\\
      In this case, cosmological defects are formed by the branes that
      have less than three spacial dimensions in the uncompactified
      spacetime.
      In the previous discussions, it was concluded that the
      cosmological production of
      monopoles and domain walls are negligible.
      In this paper, however, we reconsider the cosmological formation of
      monopoles and domain walls and show
      explicitly how they can be formed by the
      brane dynamics.
\item Defects are formed by the spatial deformation of branes.\\
      In this case, cosmological defects are formed by the continuous
      deformation of branes.
      First in ref.\cite{Alice-string}, and later in
      ref.\cite{matsuda_MQCD_defect,incidental}, the 
      fields that parameterize the positions of the branes are shown to 
      fluctuate in spatial directions to form the cosmological defects. 
      Cosmic strings are constructed in ref.\cite{Alice-string}, where
      the singularity is resolve by smearing the brane.
      Monopoles, strings and domain walls are discussed in
      \cite{matsuda_MQCD_defect, incidental}, where the relative
      positions between branes are utilized.
\item The center of the localized matter fields are shifted in the defect.\\
      In ref.\cite{Localize_on_fat_wall}, the localization of the matter
      fields on a fat domain wall is discussed to explain the small
      interactions.
      This idea is important when the suppression factor is required
      in the interaction, so that it does not violate the constraints
      from the proton lifetime.
      In ref.\cite{matsuda_shift}, we have constructed defect
      configurations, which induce the shift of the center of the
      localized matter fields. 
      These defects are used to produce the baryon
      asymmetry of the Universe\cite{matsuda_shift, Defect-baryo-4D}.
\end{itemize}

In this paper, we reconsider the cosmological formation of
brane defects, which are described by branes which have less than three
spacial dimensions in the uncompactified space.
In our model, the defect branes are extended
between branes.   
We will also consider a cosmological brane
creation induced by the excitation of a sphaleron-like brane
configuration.  

First we will review the essence of the previous arguments about the
cosmological formation of brane defects.
In the original scenario of brane inflation\cite{brane-inflation0},
inflationary expansion is driven by the potential between D-brane
and anti-D-brane evolving in the bulk.
Then the scenario of the inflating branes at a fixed angle is studied
in ref.\cite{angled-inflation}, where the slow-roll condition is 
improved by introducing a small angle.
The end of brane inflation is induced by the brane
collision where brane annihilation (or recombination) proceeds through
tachyon condensation\cite{tachyon0}.\footnote{It is possible to
construct models for brane inflation without tachyon
condensation\cite{matsuda_nontach}.}  
During brane inflation, tachyon is trapped in the false vacuum, which
may result in the formation of lower-dimensional branes after brane
inflation. 
The production of cosmological brane defects was discussed
in ref.\cite{angled-inflation, angled-defect}, where it was concluded
that cosmic strings 
are copiously produced in these scenarios, but monopoles and domain
walls are negligible. 
In ref.\cite{Majumdar_Davis}, however, it was discussed that all kinds 
of defects are produced, and the conventional problems of 
domain walls and monopoles arise.
Later in ref.\cite{RR-dvali-sting, D-brane-strings,
Halyo, BDKP-FI}, the formation of cosmological defects was reexamined 
and the conclusion was different from \cite{angled-defect} and
\cite{Majumdar_Davis}. 
As was noted in ref.\cite{angled-defect}, the effect of compactification
is significant for the defect formation, if it is induced by tachyon
condensation after brane inflation. 
Since the compactification radius must be small compared to the
horizon size during inflation, any variation of a field in the
compactified direction is suppressed.
Then the daughter brane, which is the defect formed by tachyon
condensation on the worldvolume of the mother brane, wraps the same
compactified space as the mother brane. 
As a result, the codimension of the daughter brane should lie within the
uncompactified space.
Since the number of the codimension must be even, the cosmological
defect is inevitably a cosmic string.
In ref.\cite{RR-dvali-sting}, it was discussed that the analysis
does not fully account for the effects of compactification, because the
directions transverse to the mother brane had not been considered.
In ref.\cite{RR-dvali-sting}, the effect of the RR fields, which are extended
to the compactified dimensions, were discussed.
Their conclusion was that the creation of the gradients of the RR fields
in the 
bulk of the compactified space is
costly in energy, so that the creation of the daughter D brane is
suppressed if they do not fill all the compactified dimensions.

\begin{figure}[ht]
 \begin{center}
\begin{picture}(410,300)(0,0)
\resizebox{15cm}{!}{\includegraphics{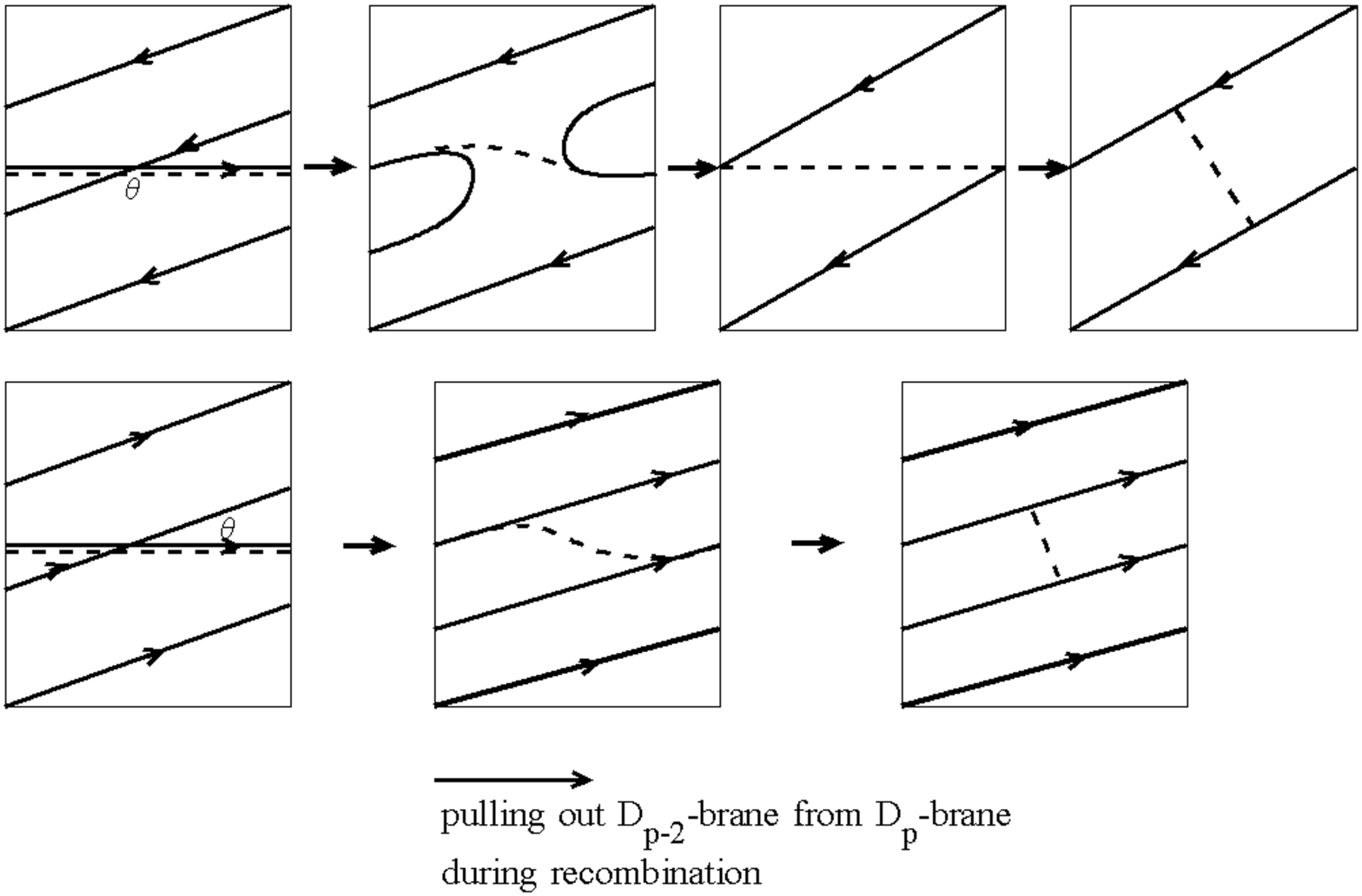}} 
\end{picture}
\caption{Upper row: schematic recombination of two $D_p$-branes with
  $(\pi-\theta) \ll 1$. The dashed line on the
  $D_p$-brane represents the $D_{p-2}$-brane that might appear on the
  worldvolume of the $D_p$-brane when the tachyon condenses.
To be more precise, a careful treatment of the effective
action shows that the eigenfunction of the tachyonic 
mode is localized on the intersection.
Since the mechanism of this localization is {\bf different} from the Kibble
mechanism, the ``seed'' for the $D_{p-2}$ brane can be localized on the
intersection. 
As the recombination proceeds, the $D_{p-2}$ brane is
pulled out from the mother brane, and finally becomes extended
between the mother brane.
In this case, the problem of the RR field is avoided since the length
of the extended 
$D_{p-2}$ brane vanishes at the time when it is pulled out from the mother
brane.
Of course, it costs energy to pull $D_{p-2}$ branes out from the mother
  branes, 
however in this case the cost is paid by the repulsive force between the
splitting mother branes.
  Second row: schematic recombination of two $D_p$-branes with 
  $\theta \ll 1$. 
  As a result, the daughter $D_{p-2}$ brane does not wrap the same
  compactified space as the mother brane.
  Thus, for the angled inflation model, the
  cosmic strings can depend on $\theta$, which is the property required
  from analysis of the effective Lagrangian.}
\label{fig:split}
 \end{center}
\end{figure}

Here it should be noted that the above arguments are not fully reliable.
The most obvious example is the inconsistency of the tension of strings
produced after angled brane inflation. 
In the original argument in ref.\cite{angled-defect}, the
$\theta$-dependence of the tension of the string did not coincide with
the one calculated from the effective Lagrangian.
What is wrong with the above arguments? 
The reason is clearly described in ref.\cite{matsuda_angleddefect}.
In the original argument of ref.\cite{angled-defect}, it was discussed
that the daughter branes are created on the mother brane.
In fact the statement is true, but we should be more careful about the
process of 
the brane recombination. 
When the distance between the mother branes becomes shorter than a
critical distance, tachyon starts to condensate on a mother brane.
Therefore the tachyon forms strings on the mother brane when the mother
branes start to recombine.
However, as we have depicted in fig.\ref{fig:split}, the daughter
$D_{p-2}$-brane must be pulled out of the $D_p$-branes.
As a result, the daughter $D_{p-2}$ brane does not wrap the same
compactified space as the mother brane.
%% added %%
The tension of the cosmic string is calculated in
ref.\cite{matsuda_angleddefect}, which depends on $\theta$ and matches
to the effective action. 
Therefore, the daughter brane in fig.\ref{fig:split} satisfies the 
property required from analysis of the effective Lagrangian.
Thus for the angled inflation model, it is obvious that the
cosmic string does not wrap the same compactified space as the mother
brane. 
From the above discussions, it is appearent that the original argument
of ref.\cite{angled-defect} 
fails because there is no obvious reason that in their final state the
daughter branes wrap the same compactified space as the mother brane.\footnote{We will not address any objection to the string formation in
angled inflation.
The defect has {\bf the same number} of
directions extended in the same compact space, which of course suggests
that the defect is a cosmic string.
The arguments in ref.\cite{angled-defect} are incorrect because they have
claimed that the daughter brane is formed on the mother brane and wraps
precisely {\bf the same compact space}  as the mother brane even in its
final configuration.
Therefore, the tension of the cosmic string calculated in
ref.\cite{angled-defect} does not match to the effective action. 
In fact, the daughter brane in angled inflation
is extended between the
recombined mother brane, which {\bf cannot wrap the same compactified
space as the mother brane}.} 
%%added%%
The idea, which supports the formation of daughter branes that are
extended between 
mother branes, is quite important for our discussions in this paper.

Since the previous arguments cannot fully account for the generic processes
of cosmological defect formation, one can hardly accept their
conclusions without reconsiding the production of monopoles and domain
walls paying careful attention to the brane dynamics. 
Is it really impossible to produce sufficient amount of cosmological
monopoles and domain walls by the branes? 
As we have discussed in ref.\cite{matsuda_MQCD_defect,
matsuda_angleddefect}, crucial modifications are required for the previous
arguments, since it is possible to produce daughter branes that
do not wrap the same compactified space as the mother brane.
In this paper we will reconsider the formation of cosmological monopoles and
domain walls by using the new idea.
For the extended branes to be produced between mother branes, 
the distance between mother branes are required to vanish at a
moment.\footnote{Our idea is generic and applicable to other conventional
cosmological processes. 
The spontaneous symmetry breaking in the effective action is sometimes
described by the recombination of the 
branes\cite{Intersecting_Brane_Models} or by the branes falling apart,
which can be induced by the thermal
effects\cite{Thermal_brane_Inflation}.
Besides the cosmological defects that are formed by a brane creation,
one may consider another type of defects that are formed by the continuous
deformation of the branes.
The two kinds of brane defects will be produced 
by the same process, as we have discussed in ref.\cite{matsuda_MQCD_defect,
incidental}. 
Therefore the analyses of the cosmological evolution of the mixture of these
defects are quite important.}
The obvious examples are;
\begin{itemize}
\item Recombination after angled brane inflation.\\ 
       At the time when the recombination starts after angled brane
      inflation, the distance between branes vanishes as required. 
       Thus, daughter branes can be pulled out of splitting branes,
       which finally become the cosmological defect in the effective action.
      As a result, the cosmological defect in the effective action is
      the brane extended between 
      mother branes. 
       Two examples are already discussed in
      ref.\cite{matsuda_angleddefect} and 
       ref.\cite{matsuda_MQCD_defect}, where the formation of cosmic
      strings are considered. 
\item Two branes coincide during inflation.\\
      In the effective action, the brane distance plays the role of a
      trigger field of hybrid inflation.
      This type of brane inflation is already discussed in
      ref.\cite{Halyo_resolvedInflation} and
      ref.\cite{Thermal_brane_Inflation}, in which the formation of
      unstable semilocal strings are suggested.
       In this paper, using the model in
      ref.\cite{Halyo_resolvedInflation}, we will show that a point-like
      object that 
       corresponds to a monopole in the effective action can
       be produced as a daughter $D3$ brane.
      The daughter $D3$ brane is extended between $D5$ branes. 
       We calculate the tension of the extended object and show that it
      coincides with the mass of a monopole in the effective action. 
\item Branes collide or oscillate with huge kinetic energy.\\
      After brane collision, there would be chaotic processes, which include
      oscillation, recombination and production/annihilation of branes.
      Therefore, it seems natural to expect that there could be a stage
      of preheating after inflation, which induce large occupation
      numbers for long-wavelength configurations that is necessary for
      the domain wall production from sphalerons.
      Then, extended objects could be formed efficiently between 
      oscillating branes. 
      As usual, these non-equilibrium processes are always important for the
      discussions about cosmological defect formation.
      As in the case of the conventional electroweak phase transition,
      we do not exclude the possibility that the final state of the 
      oscillating branes are separated at a distance, which implies a
      spontaneous symmetry breaking in the effective Lagrangian.
\end{itemize}
The defect branes could be created by the tachyon condensation on the
mother brane, or could have been existed from the beginning and have been
dissolved in the mother brane. 
In section 2, we consider the model of brane inflation in 
ref.\cite{Halyo_resolvedInflation} and study the cosmological formation
of monopoles after brane inflation.
In section 3, we discuss the formation of cosmological domain walls.
As expected, there are notable differences between the previous
arguments and our conclusions.
The cosmological formation of monopoles and domain walls are
not always negligible in models of brane cosmology.
Our results are consistent with the analyses of the effective action.

\section{Monopoles}
\begin{figure}[h]
 \begin{center}
\begin{picture}(300,200)(0,0)
\resizebox{10cm}{!}{\includegraphics{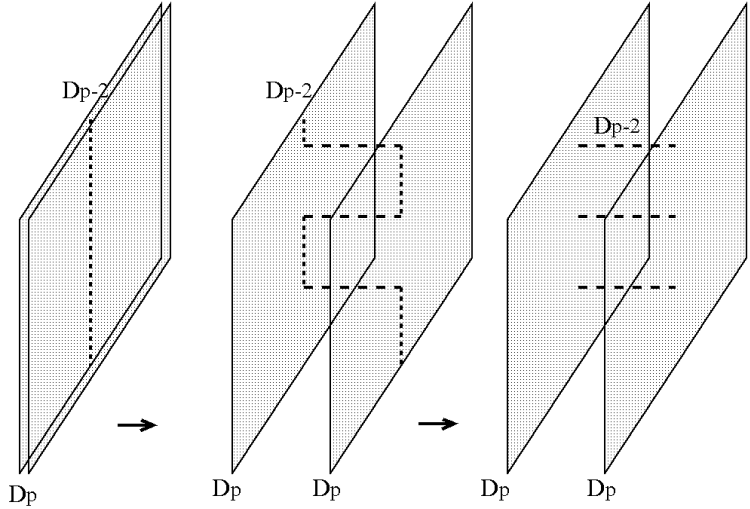}} 
\end{picture}
\caption{The initial configuration is depicted in the left picture,
  where two $D_p$ branes are located on top of each other.
  Daughter $D_{p-2}$brane is formed when tachyon condenses.
  The daughter brane is denoted by the dotted line. 
  Obviously, the previous arguments are correst so far.
  The crucial difference appears when the $D_p$ branes start to fall apart.
  The location of the $D_{p-2}$ brane fluctuates between the two
  $D_p$ branes, along the spatial
  directions of the four-dimensional space time.
  It should be noted that conventional cosmological strings are {\bf
  not} formed in this 
  case, because spacial fluctuations are inevitable.
  Thus in the final state, which is given in the right picture, the
  daughter $D_{p-2}$ brane becomes monopoles and are extended
  between the $D_p$ branes. 
  Monopoles could be connected to anti-monopoles by fluxes.
  We stress that the above mechanism works in generic
  cases of brane collision, when branes
  oscillate and produce another branes on their worldvolume. }
\label{fig:monopole}
 \end{center}
\end{figure}

In this section, we discuss the formation of monopoles in a specific
example of brane inflation.
We first repeat the previous arguments in ref.\cite{angled-defect,
RR-dvali-sting}, which suggest that monopoles are not produced 
by the usual daughter brane creation after inflation.
As is noted in ref.\cite{angled-defect}, the effect of compactification
is significant for the daughter brane creation.
Since the compactification radius is small compared to the
horizon size during inflation, any variation of a field in the
compactified direction is suppressed.
As a result, the daughter brane wraps the 
same compactified dimensions as the mother brane.
The codimension of the daughter branes should lie
within the uncompactified space, which suggests that the
defect is inevitably a cosmic string.
Seeing the above arguments, one might conclude that the monopoles are not
produced after brane inflation.
Is the production of monopoles really impossible?
To answer this question, we consider a simple example
where monopoles are produced by the daughter brane creation after brane
inflation. 
To avoid the ambiguity of the brane products, we consider a
model in which the daughter brane formation is
described without such ambiguity.
The schematic representation of our idea is given in fig.\ref{fig:monopole}.
The typical configuration of fig.\ref{fig:monopole} will also appear
in generic cosmological processes where branes collide, recombine or
stick together by thermal effects.

In any case, if the production of monopoles or domain walls is
suggested in the analysis of the effective action, it is natural to
think that it should be possible 
to explain how they can be formed by the cosmological brane dynamics.
Moreover, even if the production of such defects is not suggested in the
effective action, they might be produced by the pure brane dynamics
beyond the cut-off scale of the effective action.  
\footnote{The properties of such cosmological brane defects
might be different from the conventional defects in the effective
action. For example, Q-balls in the brane world is different
from the ones in the effective action\cite{BraneQball}, which suggests
that the cosmological defects might provide us the proof of the
underlying brane world\cite{matsuda_MQCD_defect}.}

\subsection{Brane inflation on conifold}
First we make a brief review of the Halyo's idea for brane inflation
on conifolds\cite{Halyo_resolvedInflation}.
The important aspect of this scenario is that the location of the
inflating branes coincide at the end of inflation, then the branes
separate along the $S^2$ compactified space. 
The model is described as a D-brane inflation on fractional $D3$ branes
transverse to a resolved and deformed conifold.
The D-term inflation is induced by the conifold that is resolved
by the blow-up of its tip.
In the effective action, the model looks like a model of hybrid D-term
inflation.
The slow roll is described by the slow motion of the two fractional
$D3$ branes approaching each other along a compactified direction,
and the trigger field parameterizes the  
distance between the two branes along the other compactified space 
in the base of the conifold.
For example, one can consider a conifold that is described as a cone over
$S^2\times S^3$, where a $D3$ brane, which is transverse to the conifold, 
is separated into two fractional D3 branes on the $S^3$.
The conifofd singularity is resolved by replacing the tip of the
cone with an $S^2$ of finite size, which induces inflation.
The inflaton mass arises from another deformation of the conifold
towards $ALE\times T^2$ and
results in a slow motion of the two fractional branes towards each
other.

As we are not interested in the parameter space of inflation itself,
we spare the discussions about the condition for successful
inflation.
For us, the important point is to understand the formation of daughter
branes after inflation, which finally become extended between the
separating mother branes.
As was already discussed in ref.\cite{Halyo_resolvedInflation}, 
one usually expect that the cosmic strings, which has codimension 2 in
the uncompactified space, are formed after inflation.
If the inflating $D3$ brane is a $D5$ brane wrapped on the $P^1$ with radius
$R$, the tension of the $D3$ brane is
\begin{equation}
T_{D3} =T_{D5}\int_{S^2} \sqrt{det(G+B)},
\end{equation}
where $G$ and $B$ are the $P^1$ metric and NS-NS field on the $P^1$
respectively.
Taking $B=0$ for simplicity, one can find
\begin{equation}
T_{D3}=\frac{4\pi R^2}{(2\pi)^6g_s l_s^6}.
\end{equation}
The tension of the inflating $D3$ brane must be equal to the energy density
during inflation, which corresponds to the anomalous D-term in the
effective Lagrangian.
Thus in this case, the relation $T_{D3}=g^2 \xi^2$ is required, where
$g$ is the gauge coupling in the effective action.
The cosmic strings are produced by tachyon condensation,
which are the daughter $D3$ branes on the worldvolume of the inflation brane.
The daughter brane wraps around the same compactified space as the mother
brane.
In this case, however, the symmetry of the corresponding field in the
effective action allows only semilocal strings, which is unstable to
dissolve in space, as is expected from the brane dynamics.

Let us consider a configuration of fig.\ref{fig:monopole}.
If the seed of the daughter brane fluctuate between the two coincidental
mother branes, it will be
pulled out of the mother branes.
The branes that are extended between mother branes are point-like
objects in the effective action on the mother brane. 
It is easy to calculate the tension of such
daughter branes and compare it with the mass of the monopoles in the
effective action.\footnote{It is already discussed in ref.\cite{MQCD_review}
that a $D$ string ending on a $D3$ brane provides a magnetic source for
the three-brane worldvolume gauge field.
Our example is a simple modification of the scenario.
However, it was still unclear if such extended branes are produced after
brane inflation.
As we have repeated above, the previous arguments were negative to the
cosmological production of such monopoles.}
In the final state of brane inflation,
two branes are separated at a distance of $2\pi l_s^2
\sqrt{\xi}$.\footnote{Here we follow the
definitions and notations of ref.\cite{Halyo_resolvedInflation}.}
The effective mass of the extended daughter brane is
\begin{eqnarray}
M_{ext} &=& \frac{1}{(2\pi)^4 g_s l_s^4} \times (4\pi R^2) 
\times (2\pi l_s^2 \sqrt{\xi})\nonumber\\
&=&\frac{\sqrt{\xi}}{g^2},
\end{eqnarray}
where the gauge coupling is given by the formula
\begin{equation}
\frac{1}{g^2}=\frac{R^2}{2\pi^2 g_s l_s^2}.
\end{equation}
It is easy to see that the mass of the extended daughter brane 
coincides with the mass of the monopole in the effective
action\cite{MQCD_review}. 

\begin{figure}[h]
 \begin{center}
\begin{picture}(300,200)(0,0)
\resizebox{10cm}{!}{\includegraphics{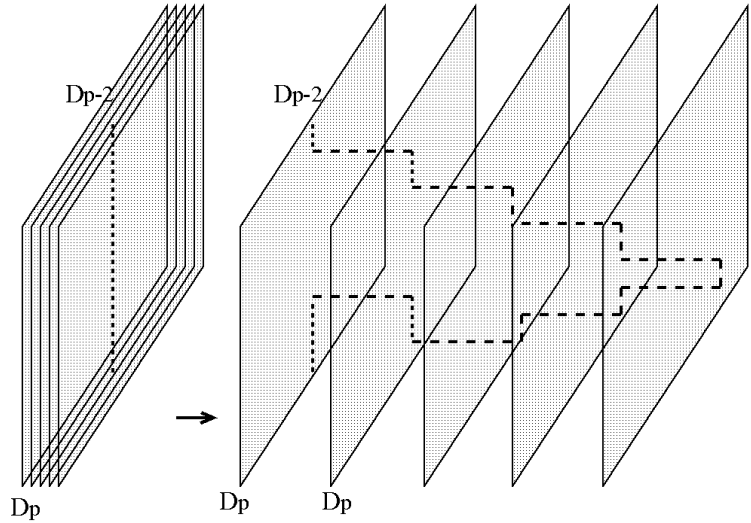}} 
\end{picture}
\caption{The initial configuration is depicted in the left picture,
  where n $D_p$ branes are located on top of each other.
  The $D_{p-2}$ brane that is created on the
  world volume of the mother $D_p$ branes is denoted by the
  dotted line.
  The location of the $D_{p-2}$ brane fluctuate in the
  uncompactified directions.
  When n branes fall apart from each other, the $D_{p-2}$ brane is
  pulled out of the mother branes. 
  In the effective action, the extended $D_{p-2}$ branes are seen as the
  point-like monopoles.}
\label{fig:monopole_N}
 \end{center}
\end{figure}

We can extend the above analysis to the cases where more than two branes
are oscillating around each other (or stick together by the thermal
effects) and then fall 
apart along a direction of the compactified space.
In this case, one can construct $N-1$ monopoles, which are the 
daughter branes extended between $N$ mother branes.
A schematic representation is given in fig.\ref{fig:monopole_N}.

\begin{figure}[h]
 \begin{center}
\begin{picture}(300,200)(0,0)
\resizebox{5cm}{!}{\includegraphics{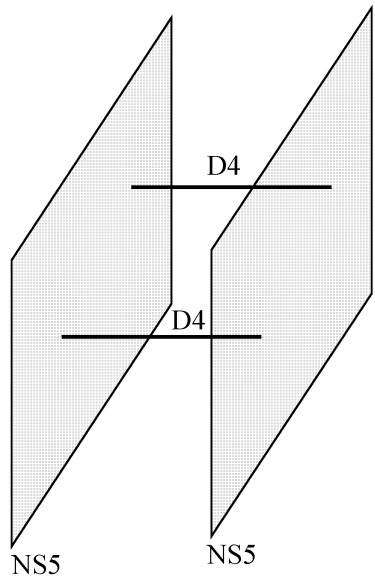}} 
\end{picture}
\caption{$N=2$ $SU(2)$ supersymmetric Yang-Mills theory realized by
  stretching two $D_4$ branes between two NS branes.
  The monopole is a rectangular $D_2$ brane with two boundaries on $D_4$
  and two on NS branes, which looks like a soap-bubble that fills the
  ``hole'' between the branes.}  
\label{fig:MQCDmonopole}
 \end{center}
\end{figure}

\begin{figure}[h]
 \begin{center}
\begin{picture}(400,200)(0,0)
\resizebox{15cm}{!}{\includegraphics{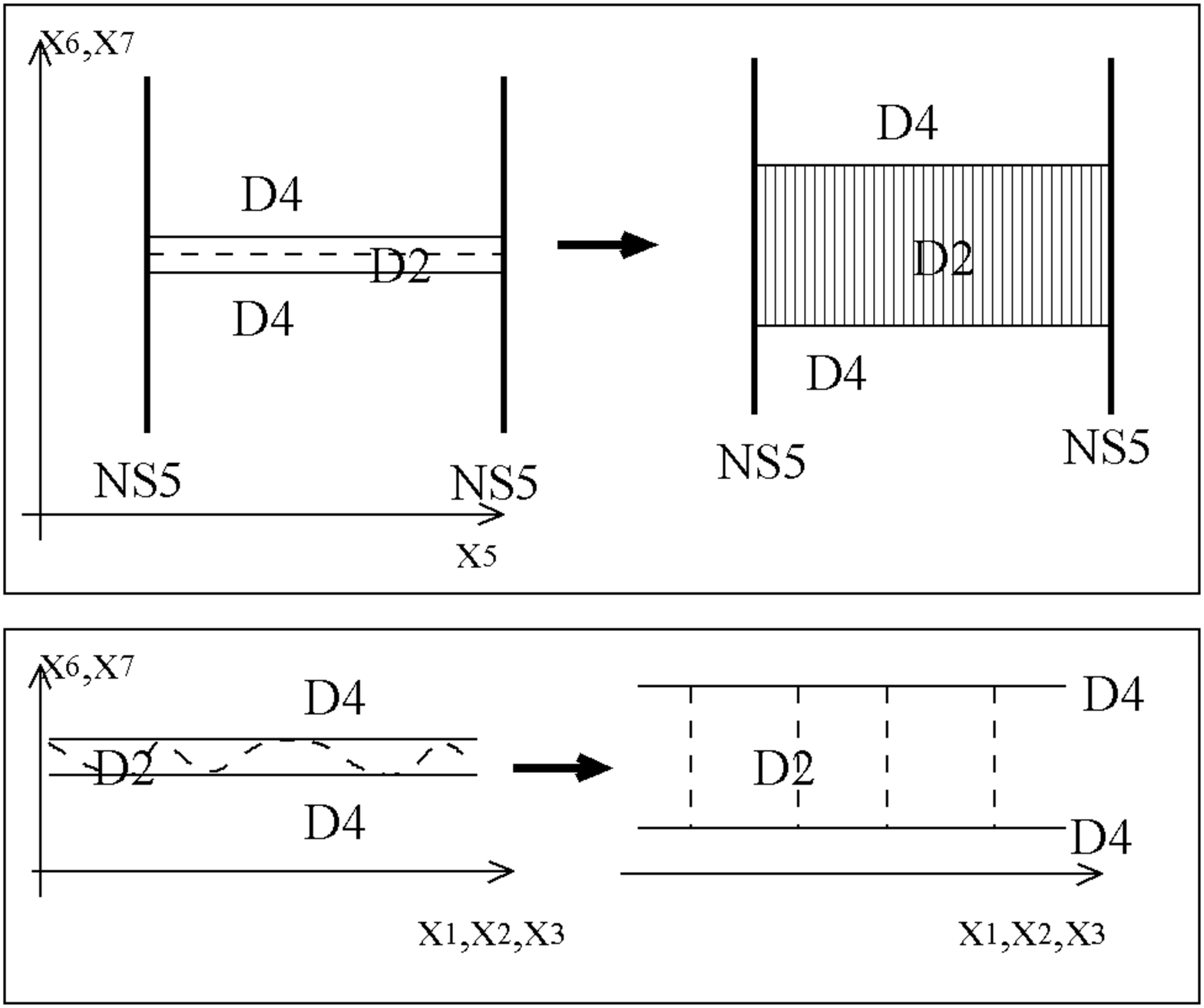}} 
\end{picture}
\caption{$D2$ brane is created on the worldvolume of a $D4$ brane
  due to tachyon condensation.
  In the initial configuration, the $D2$ brane wraps the same
  compactified space as the mother $D4$ brane.
  Thus the previous arguments are correct so far.
  However, as we have discussed, a crucial difference appears when
  the stacked mother branes start to fall apart.
  Although there is no fluctuation in the $x_5$ compactified direction,
  the $D2$ brane can fluctuate between the
  coincidental $D4$ branes along the 
  directions of the uncompactified space.
  Thus the daughter $D2$ brane is pulled out of $D4$ branes,
  which finally becomes the $D2$ brane monopoles.
  Of course our argument is consistent with a thermal history of the
  effective action if a soft mass is included.
  It should be noted that the previous arguments were inconsistent with the
  thermal history of the effective action.}
\label{fig:monopoleD2}
 \end{center}
\end{figure}

Seeing the above arguments, we cannot help thinking about the M theory
fivebrane version of QCD (MQCD), where the quantum effects are explained
by the brane dynamics.
In the most basic model of MQCD, which is depicted in
fig.\ref{fig:MQCDmonopole}, a monopole in Type IIA string
theory\cite{MQCD_monopole} is a rectangular $D2$ brane with
two boundaries on $D4$ branes and other two on NS brane.
The mass of such monopole is proportional to the minimal area of the
hole between the branes.
In realistic models, such brane configuration is expected to 
be embedded in the compactified space.
Is the cosmological formation of the $D2$ monopoles impossible?
In this case, the following conditions are required for the production
of the $D2$ brane monopoles.
\begin{itemize}
\item There was a period when the two $D4$ branes coincide.
\item $D2$ brane is created on the worldvolume of the $D4$ brane.
\end{itemize}
When the area shrinks to zero, the monopole becomes
the massless excitation.
The $D2$ brane is produced on the worldvolume of the $D4$ branes, 
due to the usual mechanism of tachyon condensation.
In fig.\ref{fig:monopoleD2}, we show why the $D2$ brane can be extended
between the separating $D4$ branes.
As usual, the daughter brane does not fluctuate in any direction of the
compactified space, while it fluctuate in the uncompactified directions.
As a result, the $D2$ monopoles are formed when the $D4$ branes start to
fall apart.
The most natural situation is that the symmetry restoration is induced
by the thermal effects, which break supersymmetry and glue branes together.
In the effective action, the restoration
of the symmetry is naturally induced by the thermal effects.
The repulsive force between the two $D4$ branes are induced by the
usual supersymmetry breaking.
The non-equilibrium production of such monopoles is possible during
brane oscillation. 
The massless monopoles gain mass when the branes start to 
fall apart. 

As a result, we conclude that the production of monopoles after
brane inflation is quite natural, even if the monopoles are the 
daughter branes created by the usual mechanism of tachyon condensation.

\section{Domain Walls}
In the previous arguments in ref.\cite{angled-defect,RR-dvali-sting},
it was concluded that the formation of cosmological domain walls
are always negligible after brane inflation, because the daughter branes must
wrap the same compactified space as the mother brane.
Is it impossible to produce daughter branes that do not wrap the
same compactified space as the mother brane?
First, we repeat the previous discussions, which suggest that the
cosmological formation of domain walls is forbidden.
As is discussed in ref.\cite{angled-defect} and \cite{RR-dvali-sting}, 
the effect of compactification is believed to play significant roles in
the defect formation due to tachyon condensation. 
Since the compactification radius should be small compared to the
horizon size during inflation, any variation of a field in the
compactified direction must be suppressed.
As a result, the daughter brane will wrap the 
same compactified space as the mother brane.
At the same time, since the codimension of the daughter branes should lie
within the uncompactified space, the
defect is inevitably a cosmic string.
On the other hand, as we have discussed in
ref.\cite{matsuda_angleddefect} for cosmic strings and in the previous
section for monopoles, it is possible to produce daughter branes that
are extended between branes, which correspond to the D-term
strings or monopoles in the effective action.
As we have seen in the previous section, efficient production
of such monopoles is allowed in generic situations.
For cosmic strings, the idea of the extended daughter brane
formation was used to solve the problem of the angle-dependence of the
string tension 
in the scenario of angled brane inflation\cite{matsuda_angleddefect}. 
The cosmological formation of the extended branes plays important
roles in the analyses of strings and monopoles.
Finally, we will examine the formation of cosmological domain walls.

Our purpose in this section is to examine the cosmological formation of 
domain walls. 
%%2 added%%
Actually, it is possible to show an example of the domain wall
formation, although this example is not a pure creation of branes
induced by the tachyon condensation on the
mother brane.

Please imagine that two coincident 5-branes are placed on top of each
other.
Both of them have four
space-time dimensions(uncompactified) and also wrap two-dimensional
compactified space.
These 5-branes look like 3-branes in the limit where the wrapped space
is very small.  
Let us assume that after a phase transition these two 5-brnes finally
fall apart to the true vacuum.

In this case, it is possible to add an additional 3-brane by hand, which
does not wrap the same compactified space as the above 5-branes.
In this case, one can 
assume that the latter 3-brane is not produced by the tachyon
condensation on the 5-brane, but could be present from the beginning. 
The latter 3-brane have four (uncompactified) space-time dimensions,
and dissolves in the 5-branes.
Then, it is quite natural to think that the additional 
3-brane sticks to either 5-brane when the 5-branes finally fall apart.
Then there appears two kinds of domains in the Universe, depending on
which 5-brane the additional 3-brane sticks to.
Since the domain walls cannot interact across the scale much larger than
the Hubble radius, one can expect that at least one domain wall is
produced in one Huble horizon.
The resulting domain wall is the 3-brane extended between the 5-branes.
%%2 added%%

%%added%%
We can discuss brane creation from brane sphalerons.
As far as the effective action is applicable,
it seems possible to create new branes that do not wrap the 
same compactified radius as the mother brane, even if it is initially
created on the worldvolume of the mother brane.\footnote{
Our argument in this section is based on the
similarity between the electroweak sphalerons and the brane sphalerons
in their 
effective action.
On the other hand, the formation of sphaleron-induced domain walls
requires some specific non-equilibrium mechanisms such as preheating or 
Parametric Resonance(PR) after inflation.
The formation of such domain walls is convinced by numerical
simulations and then it is used to advocate the new mechanism of electro
weak baryogenesis\cite{sphaleron_PR}.}
%Therefore, without simulation, it is not obvious if the
%same mechamism works in the brane-sphalerons.
%However, our aim in this paper is rather modest.
%Our aim here is not to prove the formation of domain walls, but 
%to show that the previous arguments do not cover all
%the possible mechanisms of cosmological formation of domain walls. 
%Fortunately, in section 2, we have found a suitable inflationary
%model for monopole formation. 
%On the other hand, for the formation of domain walls in section 3, 
%it is quite difficult to show that the sphaleron-induced domain walls
%are actually formed.
%The reason is apparent.
%One must first find a model of inflation in
%which the required non-equilibrium process is apparent, and second one
%must repeat the simulation if he wants to prove the formation of
%domain walls. 
%However, we cannot repeat the simulation for domain walls in this
%manuscript, because it is obviously beyond the scope of this paper.}
%%added%%

\begin{figure}[ht]
 \begin{center}
\begin{picture}(300,220)(0,0)
\resizebox{10cm}{!}{\includegraphics{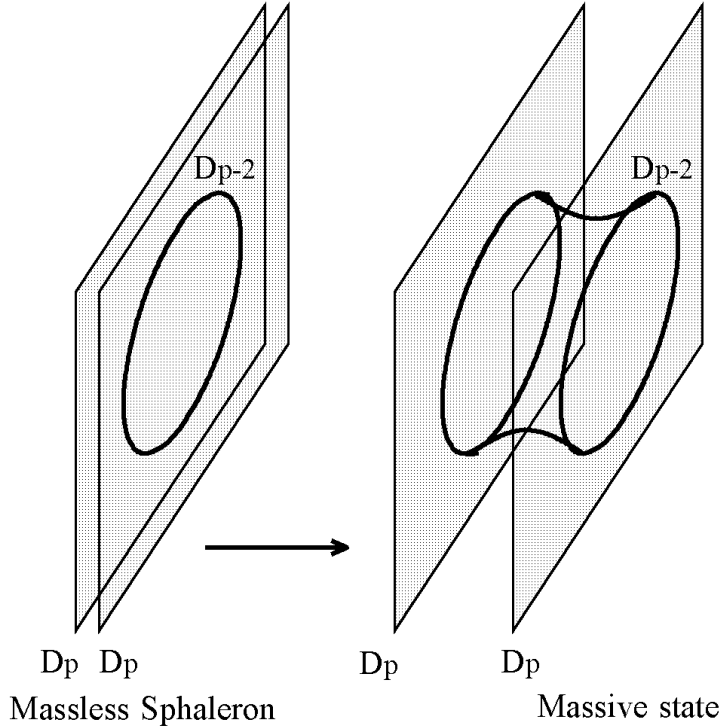}} 
\end{picture}
\caption{$D_{p-2}$ brane is extended between $D_p$ branes.
  In the left picture, the sphaleron-like brane is massless because the distance
  between the two $D_p$ branes vanishes.
  As in the cases of the conventional electroweak sphalerons, the
  production of the massless sphaleron is viable and not suppressed by the
  exponential factor.
  On the other hand, if the distance is so large that the mass of the
  sphalerons is not negligible, the excitation of sphalerons is exponentially
  suppressed.
  In our case, the produced sphaleron-like branes gains mass as the
  distance grows.
  The configuration becomes unstable if there is no other
  sphalerons in the horizon.
It might be useful to note that the sphaleron that we are considering in
this paper can be considered as the pair creation of brane anti-brane
with their boundaries on
the mother branes.
The ``throat'' solution that is constructed from  $D_{p-2}$ brane
anti-brane 
could have its boundaries on the mother $D_p$ branes.
It seems obvious that the mother branes are not necessarily the
$D_{p-2}$ brane anti-branes.
  Of course our argument is consistent with the sphaleron production in
  the effective action.}   
\label{fig:Sphaleron1}
 \end{center}
\end{figure}

\begin{figure}[h]
 \begin{center}
\begin{picture}(300,100)(0,0)
\resizebox{10cm}{!}{\includegraphics{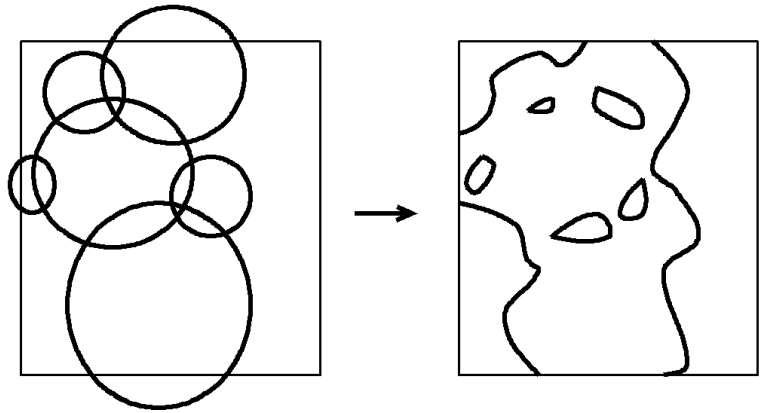}} 
\end{picture}
\caption{Sphalerons will recombine into larger (or smaller) pieces, if
  there is a viable mechanism that enhances the long-wavelength
  configuration of the sphalerons.}    
\label{fig:Sphaleron2}
 \end{center}
\end{figure}

Our second example for the formation of cosmological domain walls 
is the domain walls produced by the excitation of sphaleron-like brane
configuration\cite{Brane-Sphaleron}.\footnote{It might be useful to note that the ``throat''
configuration\cite{Brane-Sphaleron} of the $D_{p-2}$ brane could have their
boundaries on the mother $D_{p}$ branes.}
  The most obvious difference from the scenario of tachyon condensation is
that the sphaleron-like configuration 
is not required to wrap the same compactified space as the mother
brane, even when they are first created on the worldvolume of the mother brane.
In this case, for example, cosmological domain walls can be the $D3$
branes extended 
between $D5$ branes.
The schematic picture of the domain wall is given in
fig.\ref{fig:Sphaleron1}.
If there is a non-equilibrium process that enhances the
long-wavelength configurations of the sphalerons, 
the efficient production of the sphaleron excitations will happen.
Then the sphaleron-like branes will recombine, as is shown in
fig.\ref{fig:Sphaleron2}.  
Although the production of massive sphalerons is suppressed by an
exponential factor in conventional thermal processes, the production of
sphalerons are not suppressed in non-equilibrium processes such as
preheating after inflation.
Thus, the extended branes that are formed by the sphaleron-like
excitations could be produced if there was
a period when the long-wavelength configurations of the sphalerons are
enhanced.
In this case, one cannot ignore the production of domain walls.
The period when branes oscillate is inevitable in generic
cosmological models, as we have discussed in the previous section.
Since the conventional electroweak sphaleron is quite similar
to the sphaleron-like brane configuration in the effective action,
we start the discussion comparing
the electroweak sphalerons to the brane-sphalerons.

First, we briefly review the sphaleron interaction in the electroweak
gauge theory.
At zero temperature, the rate per unit volume of baryon number violating
 processes is exponentially suppressed by the factor
\begin{equation}
\Gamma(T=0) \sim exp(2S_E) \sim 10^{-170}.
\end{equation}
In the high temperature, however, the baryon number violating processes
that are mediated by sphalerons are not exponentially
suppressed\cite{electroweak_review}.
The sphalerons have typically the sizes given by the magnetic
correlation length 
\begin{equation}
\xi_\Delta \sim (\alpha T)^{-1}.
\end{equation}
In this case, one may think that the space is divided up into cells of
this size, although these massless sphalerons look nothing like
sphalerons because the ``barrier'' does not make sense when the symmetry
is restored by the thermal effects.
There is another idea for the sphaleron production\cite{sphaleron_PR},
where the non-equilibrium sphaleron transition is used to 
explain the baryon asymmetry of the Universe.

For the brane dynamics, sphaleron-like brane configuration is
discussed in ref.\cite{Brane-Sphaleron}.
The configuration is described by an appropriate excitation of the
transverse coordinate field, which corresponds to the Higgs expectation
value in the discussion of the electroweak sphalerons.
Here we do not repeat the analysis in ref.\cite{Brane-Sphaleron}, 
because the analogy comparing the sphaleron-like brane configuration to the 
electroweak sphaleron seems appropriate and sufficient for
qualitative discussions, as far as the effective action is applicable.
The sphaleron-like brane configuration is excited when the following
conditions are satisfied.
\begin{itemize}
\item More than two Branes coincide during inflation.\\
      After inflation, the branes fall apart toward the true vacuum
      configuration. 
      In the effective action, this corresponds to the symmetry
      restoration during hybrid inflation.
\item Brane inflation ends with brane collision.\\
      If the brane collision is accompanied by the brane oscillation,
      sphaleron-like branes will be produced during this
      period.
      In the effective action, this corresponds to the
      non-equilibrium sphaleron production, which is discussed in
      ref.\cite{sphaleron_PR}. 
\item Symmetry restoration is induced by the thermal effects.\\
      Spontaneous breaking of the symmetry in the effective action is
      explained by brane separation or brane recombination.
      Sphalerons are excited when they are massless.
      Sphalerons-like brane configuration is expanded between mother
      branes and obtains mass when the mother branes fall apart.
\end{itemize}
In general, one cannot simply ignore the production of the
sphaleron-like brane 
configurations, irrespective of their stability.
Even if there was no non-equilibrium processes that makes the sphalerons
form stable global structure of massive domain wall networks,
there is a possibility that unstable defects (and their decay products)
 play important roles in the evolution of the Universe. \footnote{The
domain wall that is depicted in the 
right picture in fig.\ref{fig:Sphaleron1} is unstable, because it must
shrink to a point as it obtains mass.
However, in the special cases that 
multiple sphalerons are produced at the same time within the same horizon,
multiple sphalerons will recombine into larger or 
smaller cells, as is depicted in fig.\ref{fig:Sphaleron2}.
%%2 added%%
Of course, it is not
trivial if the sphalerons could interact each other to form the
significant global structure of the domain wall networks.
If the correlation length is short, sphalerons are small even if they are
massless. 
If the nucleation rate of the sphalerons is too small
to make them interact each other to form larger structure of the domain wall
networks, sphalerons must shrink to a point after the phase transition.
In this case, sphalerons cannot become stable domain walls.
Therefore, it is required that a non-trivial mechanism that enhances the
correlation length(or the nucleation rate of the sphalerons) so that the
sphalerons interact each other to seed the domain walls.
In the scenario of preheating, the domain wall formation is already 
convinced by simulation.
In the dynamics of brane sphalerons, the brane distance is
corresponding to the higgs field.
In this respect, the domain wall formation is likely
to occur in the brane-motivated models, when some conditions are
satisfied.
As a result, it is important to note that one cannot simply ignore the
domain wall production after brane inflation.}
%%2added%%
%It is possible to construct models in which the unwanted
%sphaleron-like branes are not produced after inflation.
%For example, excitation of $D3$ branes, which are extended between a
%pair of $D4$ branes, is not allowed.
%In this case, however, another type of domain wall can be formed by the
%deformation of the $D4$ branes, not by the creation of the
%$D3$ brane. 
In ref.\cite{matsuda_MQCD_defect}, we have discussed
the formation 
of the domain walls, which are induced by the spatial deformation of the
$D4$ branes.
The domain walls formed by the spatial deformation of the branes
are, at least in principle, different from the ones that are formed by
the brane creation.
However, since the cosmological requirement for their formation is the
same\cite{matsuda_MQCD_defect}, the actual
cosmological defects are inevitably the mixture of the two 
different kinds of brane defects.

\section{Conclusions and discussions}
It was believed that monopoles and domain walls cannot be
produced by the daughter brane creation after brane inflation,
because the daughter brane must wrap the same compactified space as the
mother brane.
As we have discussed in this paper, the above statement is not fully
reliable. 
In the final state, daughter branes are not always
expected to wrap the same compactified space as the mother brane.
The most obvious example is found in the discussion about the inconsistency
of the tension of the
strings produced after angled brane inflation. 
In the original argument in ref.\cite{angled-defect}, the
$\theta$-dependence of the tension of the string did not coincide with
the the analysis of the effective Lagrangian.
In the original argument\cite{angled-defect}, it was discussed
that the daughter branes are created on the mother brane.
In fact the statement is true, but we should be more careful about the
process of the brane recombination. 
When the distance between the mother branes becomes shorter than a
critical distance, tachyon starts to condensate on a mother brane.
Therefore the tachyon forms strings on the mother brane and wraps the
same compactified space as the mother brane.
However, as we have depicted in fig.\ref{fig:split}, the daughter
brane must be pulled out if there was a recombination of the mother branes. 
As a result, the daughter brane does not wrap the same
compactified space as the mother brane.
Thus for the angled inflation model, the
cosmic strings can depend on $\theta$, which is the property required
from analysis of the effective Lagrangian.
Therefore the original argument of ref.\cite{angled-defect}
fails because there is no obvious reason that in their final state the
daughter branes wrap the same compactified space as the mother brane.  

In this paper, we have reconsidered the production of monopoles and domain
walls, paying special attention to the brane dynamics.
As a result, we have found that
monopoles are produced by daughter brane creation.
We have also suggested that domain walls could be produced by
sphaleron-like brane creation if there is an enhancement of
long-wavelength configurations of the brane-sphalerons.
Our conclusions are consistent with the analyses of the effective action.

Another type of brane defects\cite{matsuda_MQCD_defect}, which are formed
by the deformation of branes, can also be produced by the same
cosmological processes. 
Therefore, the actual cosmological relics are the
mixture of the two kinds.
Arguments about the evolution of the mixed brane defects are
interesting and deserve further investigation. 

\section{Acknowledgment}
We wish to thank K.Shima for encouragement, and our colleagues in
Tokyo University for their kind hospitality.

\newpage



\end{document}